\documentstyle[12pt,aaspp4,epsfig]{article} 

\begin{document}

\def\th{$^{13}$} 

\def\ei{$^{18}$} 

\def\tw{$^{12}$} 

\def\Lcs{{\hbox {$L_{\rm CS}$}}} 

\def\lunits{K\thinspace \kms\thinspace pc$^2$}

\def\,{\thinspace} 

\def\etal{et al.}


\def\kms{{\hbox {km\thinspace s$^{-1}$}}} 

\def\Lsun{{\hbox {L$_\odot$}}} 

\def\Msun{{\hbox {M$_\odot$}}} 

\def\Myear{{\hbox {M$_{\odot}$ y$^{-1}$}}} 

\def\ms{m\thinspace s$^{-1}$} 

\def\cm#1{{\hbox {cm$^{-#1}$}}} 

\def\cmmt{\hbox{\kern 0.20em cm$^{-3}$}} 

\def\cmmd{\hbox{\kern 0.20em cm$^{-2}$}} 

\def\Mloss{{\hbox {$\dot{M}$}}} 

\def\R*{{\hbox {$R_{*}$}}} 

\def\T*{{\hbox {$T_{*}$}}} 

\def\L*{{\hbox {$L_{*}$}}} 

\def\v*{{\hbox {$v_{*}$}}} 

\def\vt{{\hbox {$v_{t}$}}} 

\def\vp{{\hbox {$v_{p}$}}} 

\def\dv{{\hbox {$\Delta v$}}}


\font\sc=cmr10

\def\CBR{{\rm\sc CBR}} 

\def\FWHM{{\rm\sc FWHM}} 

\def\HI{{\hbox {H\,{\sc I}}}} 

\def\HII{{\hbox {H\,{\sc II}}}}


\def\Ha{H$\alpha$}                      

\def\Htwo{{\hbox {H$_2$}}} 

\def\nHtwo{n(\Htwo)} 

\def\He#1{$^#1$He}                      

\def\flecha{\rightarrow} 

\def\12CO{{\hbox {$^{12}$CO}}} 

\def\h2o18{{\hbox {H$_2^{18}$O}}} 

\def\h2o{{\hbox {H$_2$O}}} 

\def\ph2o{{\hbox {p-H$_2$O}}} 

\def\oh2o{{\hbox {o-H$_2$O}}} 

\def\n#1#2#3{{\hbox {$#1_{#2#3}$}}} 

\def\t#1#2#3#4#5#6{{\hbox {$#1_{#2#3}\!\rightarrow\!#4_{#5#6}$}}} 

\def\COJ#1#2{{\hbox {CO($J\!\!=\!#1\!\rightarrow\!#2$)}}} 

\def\CO#1#2{{\hbox {CO($#1\!\rightarrow\!#2$)}}} 

\def\CeiO#1#2{{\hbox {C$^{18}$O($#1\!\rightarrow\!#2$)}}} 

\def\CSJ#1#2{{\hbox {CS($J\!\!=\!#1\!\rightarrow\!#2$)}}} 

\def\CS#1#2{{\hbox {CS($#1\!\rightarrow\!#2$)}}} 

\def\HCNJ#1#2{{\hbox {HCN($J\!\!=\!#1\!\rightarrow\!#2$)}}} 

\def\HCN#1#2{{\hbox {HCN($#1\!\rightarrow\!#2$)}}} 

\def\HNC#1#2{{\hbox {HNC($#1\!\rightarrow\!#2$)}}} 

\def\HNCO#1#2{{\hbox {HNCO($#1\!\rightarrow\!#2$)}}} 

\def\HCOpJ#1#2{{\hbox {HCO$^+$($J\!\!=\!#1\!\rightarrow\!#2$)}}} 

\def\HCOp#1#2{{\hbox {HCO$^+$($#1\!\rightarrow\!#2$)}}} 

\def\HCOpp{{\hbox {HCO$^+$}}} 

\def\J#1#2{{\hbox {$J\!\!=\!#1\rightarrow\!#2$}}} 

\def\noJ#1#2{{\hbox {$#1\!\rightarrow\!#2$}}}


\def\Lco{{\hbox {$L_{\rm CO}$}}} 

\def\Lhcn{{\hbox {$L_{\rm HCN}$}}} 

\def\Lfir{{\hbox {$L_{\rm FIR}$}}} 

\def\Ico{{\hbox {$I_{\rm CO}$}}} 

\def\Sco{{\hbox {$S_{\rm CO}$}}} 

\def\Ihcn{{\hbox {$I_{\rm HCN}$}}}


\def\Ta{{\hbox {$T^*_a$}}} 

\def\Tmb{{\hbox {$T_{\rm MB}$}}} 

\def\Tb{{\hbox {$T_{\rm b}$}}} 

\def\Tk{{\hbox {$T_{\rm k}$}}} 

\def\nh2{{\hbox {$n({\rm H_2})$}}}

\title{Extended Far--Infrared CO Emission in the Orion OMC--1 Core$^1$}

\author {Mar\'{\i}a J. Sempere$^2$, Jos\'e Cernicharo, Bertrand Lefloch$^{3}$, 
Eduardo Gonz\'alez--Alfonso$^{4}$} 

\affil{CSIC, IEM, Dpto. F\'{\i}sica Molecular, Serrano 121, E--28006 Madrid,  
Spain.} 

\author {Sarah Leeks} 

\affil{Physics Departement, Queen Mary \& Westfield College, University
of London, Mile End Road, London E1 4NS, UK}

\footnotetext[1]{ Based on observations with ISO, 
an ESA project with instruments funded by ESA Member States 
(especially the PI countries: France, Germany, the Netherlands 
and the United Kingdom) and with participation of ISAS and NASA.} 

\footnotetext[2]{sempere@astro.iem.csic.es} 
\footnotetext[3]{Now at Observatoire de Grenoble, BP 53, F--38041 Grenoble,  
C\'EDEX 9, France. lefloch@obs.ujf-grenoble.fr  } 
\footnotetext[4]{Also at the Universidad de Alcal\'a de Henares, Departamento  
                 de F\'{\i}sica, Campus Universitario, E--28871 Alcal\'a de \ 
                 Henares, Madrid, Spain.}

\begin{abstract}

We report on sensitive far--infrared observations of $^{12}$CO  pure rotational 
transitions in the OMC--1 core of Orion. The lines were 
observed with  the Long Wavelength Spectrometer 
(LWS) in the grating mode on board the Infrared Space Observatory 
(ISO), covering 
the 43--197\,$\mu$m wavelength range. 
The transitions from $J_{up}=14$ up to $J_{up}=19$ have been identified  
across the whole OMC--1 core and lines up to $J_{up}= 43$  
have been  detected towards the central region, KL/IRc2. In addition,  we have taken  
high--quality spectra in the Fabry--Perot mode of some of the CO lines.  
In KL/IRc2 the lines are satisfactorily accounted for  
by a three--temperature model describing the plateau and ridge emission. 
The fluxes detected in the high--$J$ transitions ($J_{up} > 34$) reveal the 
presence of a very hot and dense gas component ($T=1500-2500$\,K; 
$N$(CO)$=2\times 10^{17}\cmmd$), probably originating from some of the embedded 
sources previously observed in the $\rm H_2$ near--infrared 
lines.  
At all other positions in the OMC--1 core, we estimate kinetic temperatures  
$\geq 80$\,K and as high as 150\,K at some positions around IRc2,  
from a simple Large--Velocity Gradient  
model.  
\end {abstract}


\section{Introduction} 

The Orion molecular cloud is the prime example of high--mass star 
forming regions 
and due to its proximity one of the most extensively studied sources. 
The numerous observations performed in the 
millimeter and submillimeter range (see Blake et al.\ 1987 and  
Genzel \& Stutzki 1989 for a review) have shown that the emission arises  
from three main components. 
The bulk of the molecular gas in the 
extended {\em ridge} (typical densities $\leq 10^4\cmmt$) is 
heated up to 50--60\,K by the UV field of the Trapezium stars. 
The molecular emission near the high--velocity outflow, 
the {\em plateau}, is distributed in a highly anisotropic region 
of $40\arcsec$ in size with a mean density $\leq 10^6\cmmt$. 
The plateau consists of gas heated to temperatures of 80--150\,K 
distributed in a low--velocity component centered on IRc2 and 
in a bipolar high--velocity component orthogonal to the latter. 
Finally, the {\em hot core} ($T= 400-500$\,K) of IRc2 is a compact source
(10\arcsec) with densities up to  $10^7\cmmt$.  
Its high--excitation molecular emission  
probably arises from  material shocked by the interaction of the 
high--velocity outflow with the ambient gas.

The OMC--1 core has been widely studied  in various molecular tracers. 
Among these, CO is of great interest because observations at different 
transitions delimit regions with marked contrasts in their physical 
characteristics. Most are low--$J$ ($J_{up}\leq 4$) observations that 
trace the extended low/moderate density gas, and the bulk of the 
outflow gas. 
Higher--$J$ transitions ($J_{up} \geq 6$) are sensitive to the denser and 
higher--temperature gas of the plateau or the hot core, however they are much 
less sensitive to those values from the ridge. 
Multiple CO line observations towards IRc2 carried out 
with the KAO in the submillimeter/far--infrared range (see e.g.  
Boreiko et al.\ 1989) suggest evidence for a 
hot gas component at 600--750\,K in the high--velocity 
and (possibly) low--velocity plateau, based on the few 
flux measurements at $J_{up} > 22$.  
However, the large uncertainties in the atmospheric transmission  
yield considerable variations in the parameters of this hot 
gas component.  
The ISO satellite offered us the possibility of assessing in a 
more systematic and comprehensive way the temperature distribution  
in Orion. We present in this Letter the first large--scale  
survey of the far--infrared CO lines in the Orion OMC--1 core.

\section{Observations and Results}

Five LWS--grating rasters were performed in the LWS01 mode 
(range 43--197\,$\mu$m) with a spectral resolution of 0.3--0.6\,$\mu$m 
($R=\lambda/\Delta \lambda \sim 150-300$). A total of 23 positions 
(6 scans per position) with an 
angular separation of 90$\arcsec$ ($\sim$ beam size) were observed around 
IRc2. The central position was included in each raster in order to check 
the relative calibration of the observations. The uncertainties 
in the absolute calibration are estimated to be  $\sim 30\%$.  
After pre--reduction (pipeline 7) the data were reprocessed in order to  
remove non--linear effects (Leeks et al.\ 1999).  
The size of the LWS beam is actually close to 70$\arcsec$ (E.~Caux, priv.\ comm.).

The spectra observed at each position 
(36 scans for the central one) were averaged together in order to minimize the 
noise level. The statistical errors of the measured lines fluxes are 
negligible. The transitions at $\lambda<90$\,$\mu$m are effected by additional 
uncertainties in the determination of the continuum level (hence the baseline 
subtraction), as well contamination from neighboring lines. 
The confusion level is $\sim 3\times 10^{-18}\rm\,W\cmmd$ in the averaged  
spectra. However, we checked that all the features identified in the  
averaged spectra were present in all of the original scans.  
Figure~1 shows the LWS spectra in the range 135--190\,$\mu$m for 
the positions where the CO lines could be unambiguously identified. 
Besides  C{\small II} (158\,$\mu$m) and O{\small I} 
(63\,$\mu$m and 146\,$\mu$m) fine--structure atomic lines, 
the CO emission, together with that of H$_2$O, dominates the OMC--1 far--infrared 
spectrum. Among the molecular species detected in Orion at infrared 
wavelengths, CO appears to be the most widespread (a full description of 
the IRc2 spectrum can be found in Cernicharo et al.\ 1999a, 1999b, 
hereafter C99a, C99b). 
At all of the positions, we observe the pure rotational transitions from 
$J_{up}$=14 to $J_{up}$=19, but only the lines up to $J_{up}$=17 have a 
SNR high enough to allow confident detection outside IRc2. 
At large angular distances from the central region ($> 180\arcsec$), the 
CO lines become very weak and the spectra are dominated by the atomic 
fine--structure lines. The CO integrated fluxes at fifteen 
selected positions are given in Table~1. Some of lines are contaminated by 
nearby features (O{\small I} at 145.6\,$\mu$m, OH at 163.2\,$\mu$m 
and H$_2$O at different wavelengths), whose contributions 
have been estimated from complementary Fabry--Perot observations (C99a) or 
from the grating data where the line separation was about $0.3\,\mu$m. 

Towards the central region, the consecutive pure rotational lines from 
$J_{up}$=14 up to $J_{up}$=43 could be identified. Due to the presence of 
numerous unresolved components, the calibration of the high--$J$ transitions 
($J_{up}\geq 34$) is somewhat uncertain. 
When observing the  H$_2$O lines towards IRc2 at high 
spectral resolution with the LWS in the Fabry--Perot mode ($R \sim 7000$), 
several  CO transitions, between $J_{up}$=14 and 
$J_{up}$=33, were also detected in some detector bands (Fig.~2; for details on the 
observations, see C99a). The Fabry--Perot (hereafter FP) fluxes agree to better 
than 30\% with the grating data, except for the $J_{up}$=16 and the 
$J_{up}=$28 lines where the differences reach 50\%. In view of the possible  
uncertainties in the actual pipeline, we have used the grating fluxes for 
these two lines.  
Quite remarkably, the line intensities are nearly constant from $J_{up}$=18  
up to $J_{up}$=21 indicating that the lines are 
optically thick and that the emitting regions are similar in size. 
The line intensity merely decreases by a factor of three for the following 
transitions up to $J_{up}$=28. All of the lines have typical widths of 
$\Delta v\simeq 60\,\kms$ (HPFW), i.e.\ they are 
partially resolved by the FP. 




\section{Discussion} 

\subsection{The Ridge} 

In order to determine the temperature of the gas traced by the CO  
lines detected in the OMC--1 core, we have modeled the emission of a gas  
layer with a column density of $N$(CO)$= 10^{19}\cmmd$ for various  
densities, $n({\rm H}_2)$, and temperatures.  
The CO fluxes were computed using a simple Large--Velocity Gradient approach.  
For the LWS wavelength range the dust opacity is still low enough that the  
coupling between dust and gas can be neglected.  
The adopted linewidth in the model is $\Delta v=10$\,km\,s$^{-1}$ for positions 
close to the center ($\pm 90 ''$, $\pm 90 ''$) and $\Delta v=5$\,km\,s$^{-1}$ 
for the points more distant from IRc2.

We display in Fig.~3\,b--d the expected fluxes for the transitions $J_{up}\leq 25$  
with $n({\rm H}_2)$ densities in the range $4-40 \times10^4\cmmt$ for
specific temperatures.  
The temperatures shown are those which give the best match to the fluxes  
measured at three typical positions in the core:  
halfway between the bar and the main core ($90\arcsec$, $-90\arcsec$); 
south of IRc2 towards the S6 protostellar source ($0\arcsec$,
$-90\arcsec$); the extended ridge (0$\arcsec$, -180$\arcsec$). 
For densities in the range $4-40 \times10^4\cmmt$, typical of those  
measured in the OMC--1, temperatures larger than 80\,K are required to 
account for {\em all} of the observed fluxes. Lower temperatures of $\sim
\rm 60\,K$ fail  to reproduce the observations by a factor of two or more (see Fig.~3).  
Around the central position, ($\pm 90\arcsec$, $\pm 90\arcsec$), 
we find evidence for warm gas with  $T\geq 120$\,K and  densities 
of the order of $\sim 10^{5}\cmmt$. 
 The column density adopted in our model appears as an upper limit since 
it was observed towards the IRc2 core (Blake et al.\ 1987). 
If we adopt a lower column density of $5\times 10^{18}\cmmd$ for positions 
remote from IRc2, in closer agreement with CO observations at millimeter 
wavelengths (Bally et al.\ 1987), reasonable fits are obtained by 
increasing the kinetic temperature by 10--20\%, or conversely 
increasing the density by a factor of 2--3. This does not effect our 
conclusions about the extended warm component around the hot core. 
From Fig.~3, it appears that low--$J$ observations bring very little constraints 
to the temperature determinations in our modeling. 
The kinetic temperatures derived from  KAO observations of the 
$J_{up}=7$ line ($\sim 90-100$\,K) at $100\arcsec$ resolution by  
Schmid--Burgk et~al.\ (1989) agree reasonably well for the S6 region and  
close to IRc2.

On the western side of the ridge, our modeling requires temperatures above 
100\,K or more, and/or high column densities and volume densities to account 
for the observed fluxes (see Table~1). The KAO data suggest moderate 
temperatures around $\sim 50$\,K, although it is difficult to compare both 
sets of data due to the larger beam and coarse sampling of the former. 
We note that some contamination in our data from the IRc2 
region within a possible LWS error beam cannot be excluded and could 
reduce the derived column densities and/or kinetic temperatures. Also 
the LWS beam profile is known to be rotationally asymmetric for an 
extended source and could  
account for some of the discrepancies.

\subsection{The Orion--KL/IRc2 Region} 

The emission in the far--infrared CO lines from $J_{up}$=18 to 33 can  
be reasonably explained by a simple two--temperatures model of the  
plateau region. 
First we note that a single temperature model cannot reproduce 
both the flux of the high ($J_{up}> 28$) and low--$J$ lines. 
On the other hand, the range of temperatures reported in the 
extended ridge imply that its contribution becomes negligible 
with respect to the plateau (see also Fig.~3 and below) for transitions 
above $J_{up}=18$. The physical properties of the gas in the different  
regions (in particular the plateau) were chosen as 
close as possible to those of Blake et al.\ (1987).  
The low--velocity plateau is modeled as a two--shell  
region expanding at 25\,km\,s$^{-1}$ with a 
micro--turbulent velocity of 10\,km\,s$^{-1}$.  
The line fluxes were calculated with the radiative transfer code developed  
by Gonzalez--Alfonso \& Cernicharo (1997) and were convolved with the expected 
instrumental profile (see Fig. 2) which accounts for the FP instrumental  
response (E. Caux, priv.\ comm.). We have adopted standard properties for the  
dust (silicate) grains: a typical radius of 0.1\,$\mu$m and a standard 
gas--to--dust mass ratio of 100. 
The inner region  gives rise  to the entire emission of the $J_{up}=33$ and 
$J_{up}=28$ lines, to the bulk of the emission of the $J_{up}=24$ line and 
to a significant contribution of the lower--$J$ lines. The colder gas of the  
plateau also contributes to the emission of the low--$J$ lines. 
Based on our study of the H$_2$O emission in Orion by Cernicharo et al.\  
(C99a), we assume that the density is high enough to almost 
thermalize the CO lines:  $\sim 10^7\cmmt$ for the inner region and  
$\sim 10^6\cmmt$ for the external one, close to the estimates derived by  
Blake et al.\ (1987). We use the standard CO abundance of 
$1.2\times 10^{-4}$.  
The ratio of the $J_{up}=33$ to the $J_{up}=28$ lines sets an upper limit of 
$\sim 500$\,K to the temperature of the inner region: a higher value would  
imply a lower contrast between the two lines for any CO column density. 
For a lower temperature ($\sim 350$\,K), the high column density needed to  
account for the $J_{up}=33$ line would then overestimate the  
$^{13}$CO lines, whereas they are hardly detected (we adopt an upper limit  
of 3000\,Jy). We have adopted a temperature $T=400$\,K for the inner plateau  
region.   
The resulting column density is $N$(CO)$= 10^{19}\cmmd$; the outer 
and inner radii are $3\times 10^{16}$\,cm ($4.5\arcsec$) and  
$2.2\times 10^{16}$\,cm respectively. For the external shell (outer  
radius $=7\times 10^{16}$\,cm), the temperature is $T=300$\,K and the  
column density $N$(CO)$=3.5\times 10^{18}\cmmd$. 
The assumed column densities  agree with the values quoted in the 
literature (Blake et al.\ 1987).  

The agreement with the line profiles is very satisfying taking into account 
the simple hypothesis used. The apparent high--velocity emission is fully 
reproduced by convolving with the broad--wing instrumental response (Fig.~2). 
Down to the sensitivity of these observations, we do not detect in this 
wavelength range any emission from the very high--velocity gas around IRc2. 
Comparison with all the lines observed by the LWS (either in FP or grating mode) 
proves to be satisfying up the $J_{up}=33$ transition.  
We have added to 
our modeling the contribution of the extended ridge, as a layer at 
$T= 80$\,K (see Sect.~3.3) with a column density of 
$N$(CO)$= 4\times 10^{19}\cmmd$. The resulting fit and the contributions  
of the different regions are shown in Fig.~3. 
As a whole, the emission observed towards the IRc2 core can 
be satisfactorily accounted for by two regions of the 
plateau gas at different temperatures: $\sim$300\,K and $\sim$400\,K, 
and from the ridge contributing significantly up to the  
$J_{up}=16$ line. Note that since the  
central part of the line profiles is only partially resolved by the FP, 
we cannot exclude that a part of the warm component modeled actually  
arises from the hot core (where $\Delta v\approx 5\,\kms$). 
Therefore, the emission from the $T=400$\,K component should be considered as  
the combined emission of the plateau and the hot core. 
The contribution of 
the ridge seems to be underestimated in our model. However,  
as for the adjacent positions, the contribution from an error beam could  
effect the observed fluxes. 

\subsection{The High--$J$ CO lines} 

The  higher--$J$ transitions ($J_{up} > 34$) observed with the grating mode 
are largely contaminated by other emission lines. Using our FP 
data we have estimated the contribution of the strongest adjacent lines  
(OH, H$_2$O and NH$_3$) to the CO fluxes.  
In view of the high density of spectral lines for $\lambda<90$\,$\mu$m,  
additional contamination by other weaker lines cannot be discarded (C99a,b).  
We note, however, that the lines are well above the confusion level and that  
most of the observed flux must arise from CO itself.  

After correction, all the CO lines observed either in FP or grating mode lay 
above the model of the plateau and the ridge (Fig.~3a), showing evidence for  
a hotter gas component. In the absence of other information 
on the spatial location and the kinematics of this gas, we speculate that 
such emission arises from the shocked regions of  interaction 
of the Orion outflows with the ambient medium, detected in the 
2.12\,$\mu$m lines. All sources 
of strong $\Htwo$ emission (BN, KL, PK1, PK2; Beckwith et al.\ 1983) 
are located within our 70\arcsec\ beam. 
We  modeled the high--$J$ CO emission using a Large--Velocity Gradient 
(LVG) approach with 
parameters representative of the $\Htwo$ Peak~1 region: a size of 
10$\arcsec$, a density $n(\Htwo)\sim 3\times 10^7\cmmt$ for the shocked 
emitting region, and a linewidth $\Delta v= 30\,\kms$. 
The fluxes are roughly accounted for by a gas layer at 
$T\sim 1500-2000$\,K and a column density  
$N$(CO)$\sim 6\times 10^{17}\cmmd$, implying  a shell thickness  
of $1.6\times 10^{14}$\,cm. 
These values are merely indicative 
due to the lack of spectral and spatial information, but are 
consistent with those predicted by shock models (e.g., Flower \& 
Pineau des For\^ets, 1999). 




\noindent 

{\it Acknowledgements.} 

This work has been partially supported by the Spanish DGES under  
grant PB96--0883 and by PNIE grant ESP97--1618--E. SJL acknowledges receipt
of a PPARC award. 
We thank Drs~S.~P\'erez--Mart\'{\i}nez and J.R~Goicoechea for their help in
the data reduction and Dr.~E.~Caux for providing us with the Fabry--Perot 
instrumental profile.

\setlength\unitlength{1cm} 

\clearpage

\begin{table*}

\begin{tabular}{llcccc}\hline 
\hline\noalign{\smallskip}

 Transition & 14-13& 15-14 &16-15& 17-16&\\ 

 E$_{up}$(K)   & 581 &  664   &752& 846&\\ 

{$\lambda$($\mu$m)} & 185.9 & 173.6 & 162.8 & 153.2& 

 \\\hline

\multicolumn{1}{l}{Position} &  

\multicolumn{4}{c}{Flux(10$^{-18}$ W\,cm$^{-2}$)}& 

\multicolumn{1}{c}{ T(K)} \\\hline

{(0$''$, -270$''$)}&4.9&4.6&1.7&0.22&--\\ 

{(0$''$, -180$''$)}&7.5&10&6.5&2.2&80\\ 

{(0$''$, -90$''$)}&43.&42.&40&3.2&120\\ 

{(0$''$, 90$''$)}&32&48&38&4.&120\\ 

{(0$''$, 180$''$)}&2.8&4.5&2.7&1.3&80\\ 

{(-270$''$, 0$''$)}&0.8&2.2&0.33&0.3&--\\ 

{(-180$''$, 0$''$)}&1.9&1.3&0.13&--&60\\ 

{(-90$''$, 0$''$)$^{*}$}&16&17&19&1.9&120\\ 

{(90$''$, 0$''$)}&16&18&15&3.9&100\\ 

{(180$''$, 0$''$)}&7.1&7.0&4.5&2.3&80\\ 

{(270$''$, 0$''$)}&2.7&3.0&0.64&--&--\\ 

{(-90$''$, 90$''$)$^{*}$}&16&23&25&6.2&100\\ 

{(90$''$, -90$''$)}&18&19&25&3.8&120\\ 

{(90$''$, 90$''$)}&17&18&6.7&2.5&80\\ 

{(-90$''$, -90$''$)$^{*}$}&19&29&27&--&100\\

\noalign{\smallskip} 

\hline\noalign{\smallskip} 

\end{tabular}

\vskip 1cm

Table~1~: CO line fluxes measured in the  

extended ridge of OMC--1.  

The positions for which the line fluxes could be contaminated by 

emission from IRc2 within the LWS error beam are marked with an 

asterisk (see text).\\

\end{table*}

\clearpage

\figcaption[fig1.eps]{
LWS grating spectra from the observed positions around IRc2 between 
135--197\,$\mu$m. Lower marks indicate the $^{12}$CO rotational transitions 
from $J_{up}=14$ to $J_{up}=19$. At the central position 
$^{13}$CO transitions are marked by the arrows. Some spectra have been
multiplied by a scale factor indicated at the top right of each panel.
}

\figcaption[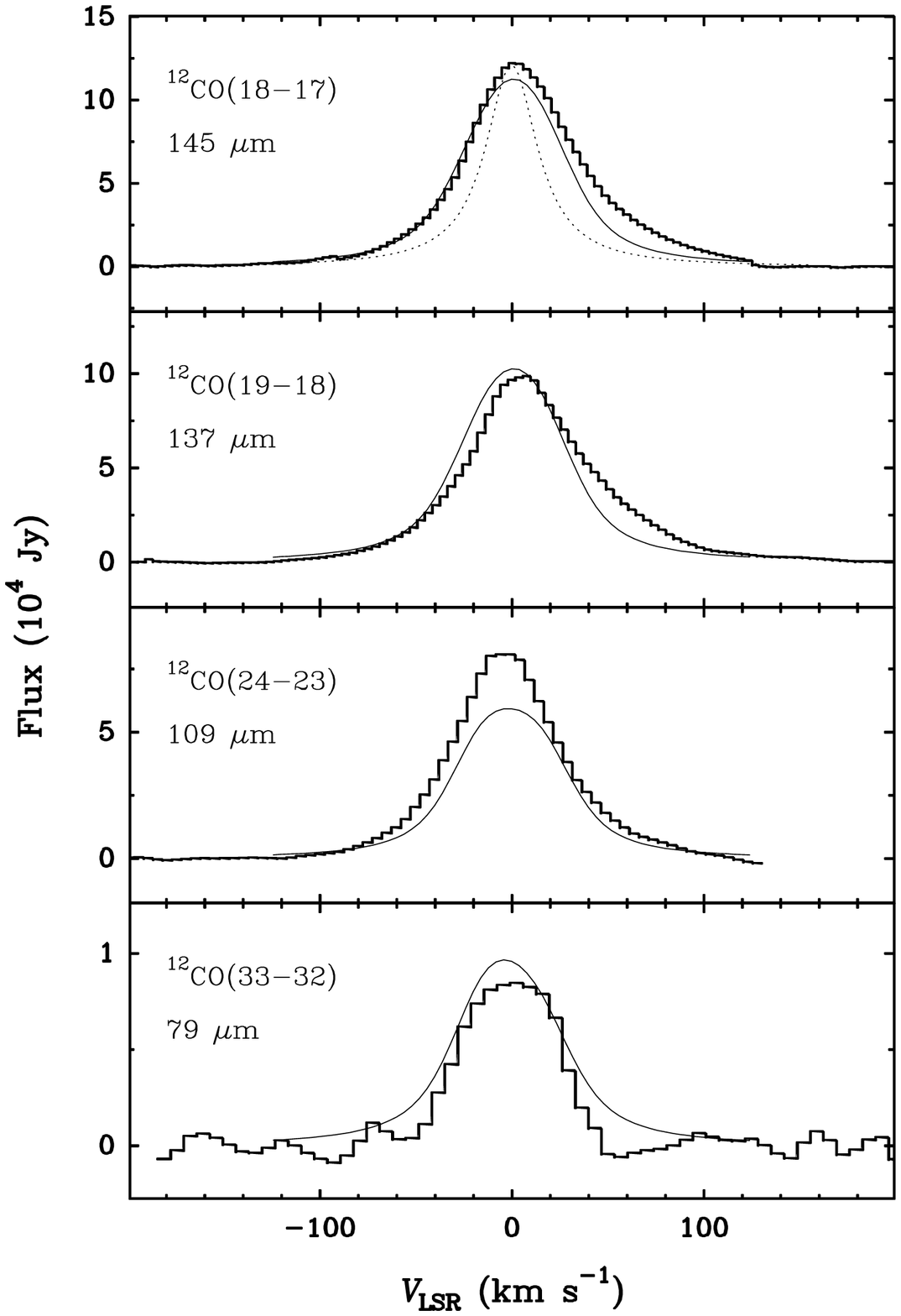]{
A montage of select CO transitions observed towards IRc2 with the LWS 
in the Fabry-Perot mode (thick line). We have indicated (thin line)  
the modeled emission of the plateau (see text). The instrumental response  
of the Fabry-Perot at $145\,\mu$m is indicated on the $J_{up}= 18$ spectra  
(dotted line).
}

\figcaption[fig3.eps]{ 
{\bf a)} Integrated flux of the CO lines detected towards the IRc2 region. 
The modeled contributions from  
the plateau (P, thin line) and the ridge (R, dotted thin) are shown. A 
hot gas component (HG)  
was introduced (dashed line) to reproduce the emission at $J_{up}>34$.  
The arrows mark the fluxes of the observed CO lines possibly overestimated  
due to contamination by weak adjacent lines.  
We have superposed (empty squares) the values  
estimated for the ISO beam size from previous work: 
$J_{up}$= 4 (Schultz et al., 1992); 6 (Graf et al., 1990); 
7 (Schmid-Burgk et al.\ 1989; Howe et al.\ 1993); 21 (Boreiko \& Betz, 1989); 
22-34 (Watson et al 1985). 
{\bf b,c,d)} Expected fluxes for the CO lines up to $J_{up}= 25$ 
observed in the extended ridge with the ISO LWS beam, for kinetic  
temperatures comprised between 60--120\,K, 
$\Htwo$ densities $4-40\times 10^4\cmmt$  and a column density  
$N$(CO)= $10^{19}\cmmd$.  We show the ``best temperature'' fit, for 
each pointing, which  
accounts for the measured fluxes (Table~1). The arrows have  
the same meaning as above.
} 

\end{document}